\documentclass[tbtags,12pt,a4paper]{article}
\usepackage{a4wide}
\usepackage{fancyhdr}
\usepackage{graphicx}
\usepackage{float}
\usepackage{enumerate}
\usepackage{amsmath}
\usepackage{amssymb}
\usepackage{amsfonts}
\usepackage{amsthm}
\usepackage{amssymb}
\usepackage{mathrsfs}
\usepackage{equation}

\begin{document}
\renewcommand{\thefootnote}{\fnsymbol{footnote}}
\def\gsim{\:\raisebox{-0.5ex}{$\stackrel{\textstyle>}{\sim}$}\:}
\def\lsim{\:\raisebox{-0.5ex}{$\stackrel{\textstyle<}{\sim}$}\:}

\begin{titlepage}

\begin{flushright}
RIKEN--TH 86 \\[-0.1cm]
October 2006\\[5mm]
\end{flushright}

\vspace{0.5cm}

\begin{center}
{\Large \bf QCD Effects in the Decays of TeV Black Holes}\\[1.cm]
   {\large Christian Alig$^{1}$, Manuel Drees$^1$, and Kin-ya Oda$^2$}
\end{center}

\vskip 0.5cm

{\small
\begin{center}

$^1$ {\it Physikalisches Institut, Universit\"at Bonn, Nussallee 12, D53115
          Bonn, Germany} \\[2mm]
$^2$ {\it Theoretical Physics Laboratory, RIKEN, Saitama 351-0198, Japan}

\end{center}
}

\vspace{1.cm}
\begin{abstract}

\noindent In models with ``large'' and/or warped extra dimensions, the
higher--dimensional Planck scale may be as low as a TeV. In that case black
holes with masses of a few TeV are expected to be produced copiously in
multi--TeV collisions, in particular at the LHC. These black holes decay
through Hawking radiation into typically ${\cal O}(20)$ Standard Model
particles. Most of these particles would be strongly interacting. Naively this
would lead to a final state containing 10 or so hadronic jets. However, it
has been argued that the density of strongly interacting particles would be so
large that they thermalize, forming a ``chromosphere'' rather than
well--defined jets. In order to investigate this, we perform a QCD simulation
which includes parton--parton scattering in addition to parton showering. We
find the effects of parton scattering to remain small for all cases we
studied, leading to the conclusion that the decays of black holes with masses
within the reach of the LHC will {\em not} lead to the formation of
chromospheres.

\end{abstract}
\vskip 0.5cm

\end{titlepage}

\section{Introduction}

The Standard Model (SM) of particle physics is very successful in reproducing
experimental data. However, from the theoretical point of view it has several
unsatisfactory features. Its perhaps biggest problem stems from the very large
hierarchy between the electroweak mass scale $M_{\rm weak} \sim 100$ GeV and
the (reduced) Planck mass $M_{\rm P} \simeq 2.4 \cdot 10^{18}$ GeV. The origin
of this hierarchy has no explanation within the SM; worse, it tends to be
destroyed by quantum corrections.

In the late 1990's it has been suggested \cite{Arkani-Hamed:1998rs,
  Randall:1999ee} that this problem could be solved by introducing additional
spatial dimensions. They could be relatively large if only gravity is allowed
to propagate in them, i.e. if all (other) SM fields are bound to a brane with
three spatial dimensions. The strength of the gravitational interactions at
lengths larger than the radii of these additional dimensions would then be
diluted by a factor which is essentially given by the ratio of the volume of
these extra dimensions and the appropriate power of the higher--dimensional
Planck length. This ratio can be very large even if the higher--dimensional
Planck mass $M_D$ is not far above $M_{\rm weak}$. If this idea is correct,
the very large measured value of $M_{\rm P}$ is an artefact of using a
3--dimensional description of a world that actually has $3+n$ spatial
dimensions.

This scenario leads to dramatic predictions for collisions of pointlike
particles at high center--of--mass (cms) energy. Collisions at energies around
$M_D$ would likely be dominated by the exchange of gravitons
\cite{'tHooft:1987rb}, rather than by exchange of SM gauge bosons. Collisions
at energies exceeding $M_D$ could lead to the formation of black holes
\cite{banks, Giddings:2001bu, Dimopoulos:2001hw, Eardley:2002re,
  Solodukhin:2002ui, Hsu:2002bd, Yoshino:2002tx, Yoshino:2005hi}, with cross
section being given by the square of the (generalized) Schwarzschild radius.
Since this radius actually increases with increasing cms energy, this would
lead to ``the end of short--distance physics'' \cite{Giddings:2001bu}.

These black holes with masses of a few TeV should decay very quickly via the
higher--dimension analogue \cite{Myers:1986un} of Hawking radiation
\cite{Hawking:1974sw}, into final states on the brane consisting mostly of SM
degrees of freedom \cite{Emparan:2000rs}. Since the Hawking temperature of
black holes that can be produced at the LHC is ${\cal O}(100 \ {\rm GeV})$,
its decay should produce roughly ${\cal O}(20)$ particles with average energy
$\sim 3$ times the Hawking temperature. This decay would appear to be
instantaneous to LHC experiments, since the typical lifetime of such a black
hole is only ${\cal O}(10^{-27} \ {\rm s})$.

Since these black holes decay into all SM degrees of freedom with
approximately equal probability, most final state particles would be strongly
interacting quarks or gluons. This leads to the expectation that a typical
black hole event at the LHC would contain 10 or so jets plus a few (charged or
neutral) leptons and/or photons, each with typical energy of several hundred
GeV. This is obviously a very dramatic signature, which should be easy to
detect.

However, it has been argued by Anchordoqui and Goldberg
\cite{Anchordoqui:2002cp} that the density of strongly interacting partons
just after the decay of the black hole would be so high that they would
frequently interact with each other, leading to the formation of a (more or
less) thermalized ``chromosphere'', i.e. a (nearly) spherical shell of
thousands of particles with rather small energies. This final state would also
be very easy to detect. However, in this case the final state would be
characterized almost uniquely by the mass of the black hole. More detailed
investigations of the primary decay spectrum, which could give information
about the number of additional dimensions as well as the spin of the decaying
black hole, could then only be performed with the (rather few) primary charged
leptons and photons.

It is therefore of some importance to decide whether the assertion of
ref.\cite{Anchordoqui:2002cp} is in fact correct. In this paper we report
results of a Monte Carlo simulation of the QCD effects relevant for the decay
of TeV black holes, including both parton showering and partonic collisions.
This simulation has to keep track of the space--time evolution of the black
hole decay products. In contrast, the usual shower codes only keep track of
the virtuality of the partons in the shower, but treat the shower itself as
instantaneous. This is quite adequate for most applications, since a parton
shower only lasts $10^{-23}$ seconds or so, corresponding to a spatial
extension of a few Fermi, many orders of magnitude below the resolution of any
conceivable detector. However, if a chromosphere forms at all, it should do so
during those $10^{-23}$ seconds. A careful treatment of the spatio--temporal
evolution of the shower is therefore mandatory for us.

We find that the effects of parton--parton scatterings after black hole decay
are essentially negligible for a black hole of 5 TeV, and are quite small even
for 10 TeV black holes. Even in the latter case, most partons do not scatter.
Moreover, most of these (relatively rare) interactions are rather soft, i.e.
they do not change the energies, trajectories or virtualities of the
participating partons very much. We therefore conclude that the decays of
black holes that might be produced at the LHC will {\em not} lead to the
formation of a chromosphere. Black hole event generators that ignore
interactions between black hole decay products \cite{charybdis,catfish} only
make a small mistake.

The remainder of this note is organized as follows. In the next Section we
describe the production of TeV black holes, and their subsequent decay through
Hawking radiation, in slightly more detail. This determines the initial
set--up of the partonic system, which later may or may not develop into a
chromosphere. In Sec.~3 we summarize the argument in favor of chromosphere
formation, closely following ref.\cite{Anchordoqui:2002cp}; we also point out
some weaknesses in this argument. Sec.~4 is devoted to a description of the
simulation program we wrote. In Sec.~5 we present numerical results for the
angular correlation between pairs of charged particles, for the overall energy
flow of the hadronic black hole decay products, and for the microscopic
structure of these events. Finally, Sec.~6 contains a brief summary and
outlook.

\section{Black hole production and decay}

Lower bounds on the classical gravitational production cross section of black
holes are obtained in~\cite{Eardley:2002re,Yoshino:2002tx,Yoshino:2005hi}.
Quantum corrections are estimated in~\cite{Solodukhin:2002ui,Hsu:2002bd}.
The implication of the correspondence principle for black holes and strings is
considered in~\cite{Dimopoulos:2001qe}. All these results suggest that the
black hole production cross section grows geometrically above the
(higher--dimensional) Planck scale:
\begin{align} \label{cs}
\sigma(\hat{s}) &\simeq \pi r_h(M)^2 \propto \hat{s}^{1/(1+n)} \, .
\end{align}
Here, $n$ is the number of additional spatial dimensions, $\hat{s}$ is the
partonic cms energy, and $r_h$ is the horizon radius of the
$D=4+n$~dimensional Schwarz\-schild black hole with mass
$M=\sqrt{\hat{s}}$:\footnote{We use the notation of \cite{Anchordoqui:2002cp},
  where $M_D$ stands for the reduced higher--dimensional Planck mass.}
\begin{equation} \label{radius}
r_h(M) = {1 \over M_D} \left[ \frac {M} {M_D} \frac {2^n \pi^{(n-3)/2} 
\Gamma\left({3+n\over2}\right) } {n+2} \right]^{1/(1+n)} \, .
\end{equation}
Eq.(\ref{cs}) should hold as long as $r_h$ is small compared to the size of
the additional dimensions, which is true for all cases of interest to LHC
experiments. 

Astrophysical processes lead to the very strong lower bound $M_D \gg 10$ TeV
for $n \leq 3$ \cite{raffelt}. For larger $n$ the lower bound on $M_D$ comes
from searches for the production of gravitons (including their Kaluza--Klein
towers) at colliders, yielding $M_D \geq 0.65$ TeV for $n=6$ \cite{pdg}.  It
has been argued \cite{anchor3} that the non--observation of black holes
produced by very energetic cosmic ray neutrinos yields the slightly stronger
bound $M_D \gsim 1$ TeV for $n \geq 5$. However, this bound relies on
assumptions on the flux of very energetic cosmic neutrinos. We will see
shortly that chromosphere formation is most likely for the smallest allowed
value of $M_D$. To be conservative, we will therefore present numerical
results for $n=6$ and $M_D = 0.65$ TeV. In that case according to
Eqs.(\ref{cs}), (\ref{radius}) the LHC operating at a proton--proton cms
energy of 14 TeV yields cross sections in excess of 200 pb (10 fb) for $M > 5
\ (10)$ TeV~\cite{Giddings:2001bu}. In other words, the LHC will be a
veritable black hole factory if a TeV scale gravity scenario is employed by
nature~\cite{Giddings:2001bu,Dimopoulos:2001hw}.\footnote{QCD initial state
  radiation is, as usual, taken into account by using scale--dependent parton
  distribution functions, the relevant momentum scale being set by $1/r_h$
  \cite{Giddings:2001bu}. Numerical simulations indicate \cite{Yoshino:2002tx,
    Yoshino:2005hi} that several tens of percent of the cms energy of the
  colliding partons may escape in form of gravitational radiation. The same
  calculations show that Eqs.(\ref{cs}), (\ref{radius}) underestimate the
  cross section for black hole production for fixed $\hat{s}$. Nevertheless
  the energy ``lost'' in gravitational radiation implies $M < \sqrt{\hat{s}}$,
  in which case the cross section for the production of black holes with a
  given mass at the LHC might be more than three orders of magnitude smaller
  than indicated by Eqs.(\ref{cs}), (\ref{radius}) with $\sqrt{\hat{s}} = M$
  \cite{anchor2}. Finally, if a ``generalized uncertainty principle'' imposes
  a lower bound on physical lengths of order $1/M_D$, the cross section for
  producing black holes with mass $M \gg M_D$ at the LHC would also be reduced
  by several orders of magnitude \cite{hoss1}.}
  
We note that the black hole is generically produced with a sizable angular
momentum~\cite{Ida1}. However, here we only consider the case of vanishing
angular momentum. While the spin of the black hole would affect the details of
its decay spectrum during the early ``spin--down'' phase \cite{Ida1, kanti,
  Ida:2006tf}, these details are not likely to significantly change the
importance of the QCD processes which are the main focus of our analysis.

Once produced, the black hole radiates off its mass~$M$ via Hawking
radiation~\cite{Hawking:1974sw} mainly into the brane--localized standard model
particles~\cite{Emparan:2000rs}.\footnote{Possible enhancement effects of bulk
  graviton emission, especially for highly rotating black holes, have been
  discussed in \cite{enhance}.} The number spectrum for the emission of a
spin$-s$ particle with energy $\omega$ per unit time is
\begin{align} \label{spectrum}
{d \dot{N}_s\over d\omega} &= {1\over2\pi} {\Gamma_s\over e^{\omega/T} -
  (-1)^{2s}}\, . 
\end{align}
Here the Hawking temperature $T$ is given by
\begin{align} \label{temp}
T &= {1+n\over 4\pi r_h}\, .
\end{align}
The ``greybody factor'' $\Gamma_s(E)$ in Eq.(\ref{spectrum}) is defined to be
the absorption rate of the incoming flux with energy $E$ at spatial infinity:
\begin{align}  \label{gb}
\Gamma_s &= {\dot{N}_\text{in} - \dot{N}_\text{out} \over \dot{N}_\text{in}},
\end{align}
when the purely in--going boundary condition is put at the black hole horizon.
Physically $\Gamma_s$ is, in the ``time-reversed'' sense, the proportion of
the radiation that passes through the gravitational potential well from the
horizon towards spatial infinity.

\begin{figure}[H] \label{specfig}
\begin{center}
\includegraphics[scale=2]{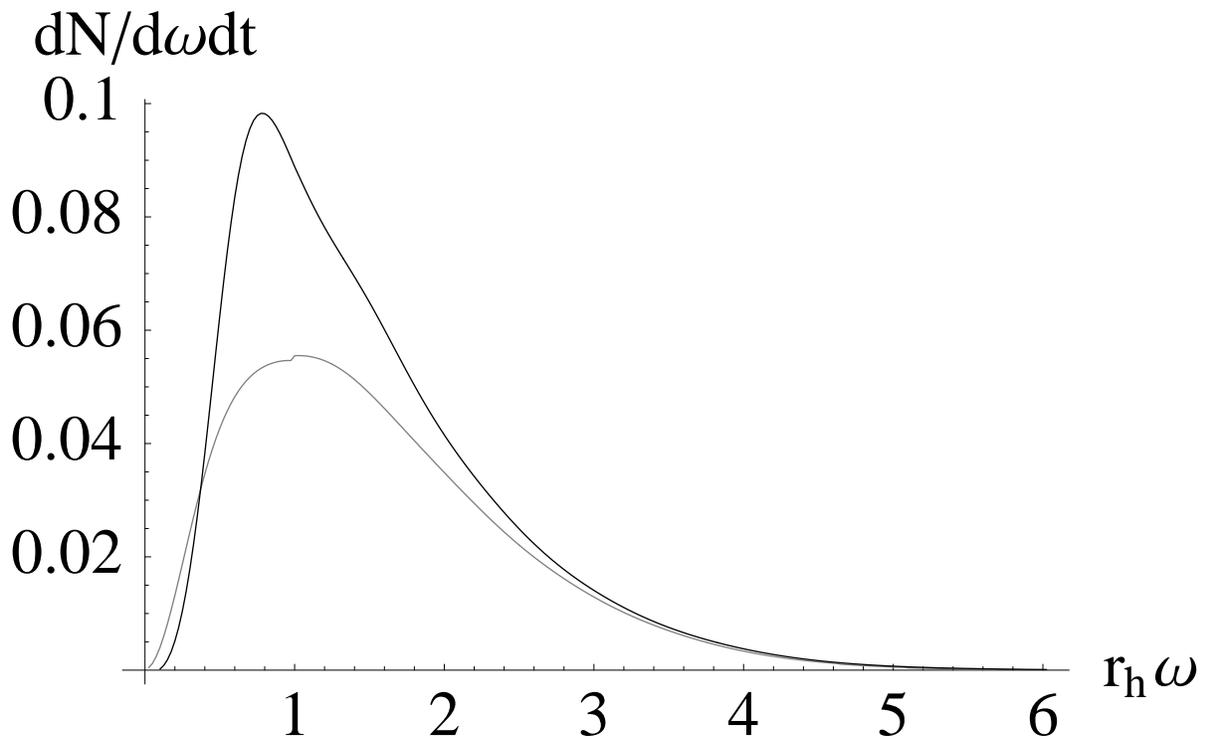}
\end{center}
\caption{The energy spectrum of gluons (upper, black) and quarks (lower, grey)
  emitted in black hole decay for $n=6$ additional dimensions. $\omega$ is the
  energy of the emitted particle, and the Schwarzschild radius $r_h$ is given
  by Eq.(\ref{radius}).}
\end{figure}

\setcounter{footnote}{0}
Numerical results \cite{Ida:2006tf} for the spectra of quarks and gluons
produced in the decay of a black hole are shown in Fig.~1. In principle
Eqs.(\ref{spectrum}) and (\ref{temp}) predict that the spectrum of black hole
decay products changes with time: as more energy is radiated off, the mass of
the black hole becomes smaller. Eq.(\ref{radius}) shows that this also reduces
its radius, which according to Eq.(\ref{temp}) increases its
temperature.\footnote{Since Fig.~1 effectively shows the spectrum in units of
  the inverse black hole radius, it includes all the information required for
  including this effect in the simulation. However, these semi--classical
  results only hold for $M \gg \omega$; they may therefore not describe the
  late stages of black hole decay adequately. In fact, it has been argued
  \cite{remnant} that a black hole will not evaporate completely, but leave
  behind a stable or at least long--lived, and possibly charged, remnant with
  mass $\sim M_D$. We do not consider this possibility in our work.} For
simplicity we employ the ``sudden decay approximation''
\cite{Dimopoulos:2001hw} where the entire decay spectrum is calculated using a
fixed black hole mass.

All degrees of freedom with a given spin are emitted with equal probability in
black hole decay; however, Fig.~1 shows that the greybody factors do depend on
the spin. In order to include this effect, we define the integrals over the
decay spectra
\begin{equation} \label{is}
I_s = \int_0^\infty \frac {d \dot{N}_s} {d \omega} d z \, ,
\end{equation}
where $z = \omega r_h$ is the dimensionless variable shown in Fig.~1.
Numerically, $I_{1/2} \simeq 0.115, \, I_1 = 0.155$. In our simulation we are
primarily interested in the question whether quarks and gluons produced in
black hole decay will thermalize. Since scattering cross sections for heavy
quarks are smaller than those for massless quarks, we conservatively assume
that only $u,d,s,c$ quarks and gluons are emitted in black hole decays. The
relative abundance of gluons is then on average given by
\begin{equation} \label{gluons}
\frac {\langle N_g \rangle} {\langle N_g \rangle + \langle N_q \rangle} =
\frac {8 I_1} {8 I_1 + 24 I_{1/2}} \, . 
\end{equation}
An analogous argument shows that about 75\% of the black hole mass will be
radiated into quarks and gluons.\footnote{Massive gauge and Higgs bosons
  produced in black hole decay frequently also decay into quarks. However,
  these bosons have much longer lifetimes than the black hole itself. The
  quarks produced in their decays therefore come too late to contribute to the
  formation of a chromosphere, although they might get trapped in the
  chromosphere \cite{Anchordoqui:2002cp} should it indeed form.}

The total (average) number of particles produced in the decay of a single
black hole can be estimated from the integrals
\begin{equation} \label{its}
\tilde I_s = \int_0^\infty z \frac {d \dot{N}_s} {d \omega} d z \, ;
\end{equation}
numerically, $\tilde I_{1/2} = 0.181, \, \tilde I_1 = 0.231$. The average
energies (in units of $r_h^{-1}$) of the produced particles are given by the
ratios $\tilde I_s / I_s$. Numerically, the average energy of a quark or gluon
is $\langle E_q \rangle = 313 \ (283)$ GeV and $\langle E_g \rangle = 296 \ 
(268)$ GeV for $M = 5 \ (10)$ TeV. Using the fact that about 75\% of the total
energy goes into hadrons, this gives average parton multiplicities
\begin{eqnarray} \label{nav}
\langle N_q \rangle &=& 8.4 \ (18.6) \nonumber \\
\langle N_g \rangle &=& 3.8 \ (8.3) \, ,
\end{eqnarray}
for $M = 5 \ (10)$ TeV, where $N_q$ counts both quarks and antiquarks.
Choosing a larger higher--dimensional Planck scale $M_D$ would lead to higher
Hawking temperature $T$, and hence to fewer, more energetic, primary black
hole decay products; this would obviously reduce the probability of
parton--parton scattering.

Finally, Eq.(\ref{spectrum}) also allows to compute the lifetime of the black
hole. To that end, one computes the time derivative of the mass of the black
hole, 
\begin{equation} \label{dmdt}
\frac {dM} {dt} = \int_0^\infty \omega \frac {d \dot{N} } {d \omega} d
\omega\, .
\end{equation}
If we set $t=0$ for the time of black hole production, its lifetime $\tau_{\rm
  bh}$ is defined by $M(\tau_{\rm bh})$ = 0. Eq.(\ref{spectrum}) shows that
the rhs of Eq.(\ref{dmdt}) is $\propto r_h^{-2}$. The use of Eq.(\ref{radius})
then leads to \cite{Giddings:2001bu}
\begin{equation} \label{life}
\tau_{\rm bh} = \Gamma_{\rm bh}^{-1} = \frac {C} {M_D} \left(\frac {M} {M_D}
\right)^{(n+3)/(n+1)}\, .
\end{equation}
The coefficient $C$ can be computed by numerically integrating the spectra
shown in Fig.~1, leading to $C \simeq 0.2$. Eq.(\ref{life}) then
gives $\Gamma_{\rm bh} = 240\ (100)$ GeV for $M= 5 \ (10)$ TeV and $M_D = 0.65$
TeV, corresponding to lifetimes of order $10^{-27}$ seconds.

Having presented the relevant properties of the partonic final state that
emerges from black hole decay, we are now ready to discuss whether
interactions between these partons might lead to formation of a chromosphere.

\section{Arguments for and against chromosphere formation}

The argument by Anchordoqui and Goldberg \cite{Anchordoqui:2002cp} starts from
the observation that the parton number density $n$ after black hole decay is
quite high: roughly ${\cal O}(10)$ partons are distributed over a sphere with
radius $\sim c \tau_{bh}$. They estimate the rate $\Gamma$ of bremsstrahlung
reactions (which increase the total number of partons) as $\Gamma = n c
\sigma_b$, with 
\begin{equation} \label{brems}
\sigma_b \simeq \frac {8 \alpha_s^3} {\Lambda^2} \ln \left( \frac {2Q}
  {\Lambda} \right) \, ,
\end{equation}
where $\alpha_s$ is the strong coupling constant and $Q$ the initial energy of
the parton in the rest frame of the black hole; the energy scale $\Lambda$ is
estimated as the inverse of the radius of the expanding shell of partons. This
leads to a total interaction rate per parton \cite{Anchordoqui:2002cp}
\begin{equation} \label{intrate}
{\cal N}_{\rm int} \simeq 0.15 \frac {N_{q,{\rm init}}} {10} \left( \frac
  {\alpha_s(Q_{\rm min}) } {0.2} \right)^3 \ln \left( \frac {2 Q} {Q_{\rm min}}
  \right) \ln \left( \frac {\Gamma_{\rm bh}} {Q_{\rm min}} \right) \, .
\end{equation}
For $Q \simeq 400$ GeV and initial parton number $N_{q,{\rm init}}=10$
Eq.(\ref{intrate}) predicts about 3 interactions per parton even with minimal
momentum transfer $Q_{\rm min} = 9$ GeV; this increases to $\sim 30$
interactions for $Q_{\rm min} =1.8$ GeV (the mass of the $\tau$
lepton). Anchordoqui and Goldberg argue that partons that interact so
frequently must thermalize, leading to the formation of an expanding shell of
particles with approximately thermal energy distribution.\footnote{In
  contrast, the authors of ref.\cite{Giddings:2001bu} state that the shell of
  black hole decay products is too thin to thermalize; however, no
  quantitative estimate of the effects of interactions between these decay
  products is given there.}

This argument has several weaknesses. To begin with, most QCD interactions are
rather soft, as seen by the factor $1/\Lambda^2$ in Eq.(\ref{brems}).
Similarly, most emitted gluons will be soft and/or collinear; this leads to
the logarithmic factor in the cross section (\ref{brems}). In other words, a
``typical'' interaction may not change the energies and momenta of the
participating partons very much, and may only lead to the emission of a soft
and/or collinear gluon. Such interactions would {\em not} impede the formation
of well--defined hadronic jets. In fact, during a QCD parton cascade, many
gluons will in any case be emitted from the partons produced in a given
``hard'' process; most of these gluons will also be soft and/or collinear.
This gives rise to finite widths and masses of hadronic jets, but does not
destroy them.

Secondly, Anchordoqui and Goldberg seem to have overlooked the fact that a
scattering reaction takes a finite time: according to the uncertainty
principle, the time at which a reaction with energy exchange of ${\cal
  O}(\Lambda)$ occurs can only be determined with an intrinsic uncertainty of
order ${\cal O}(1/\Lambda)$. Most reactions have $\Lambda \sim r^{-1}$
\cite{Anchordoqui:2002cp}, $r$ being the radius of the expanding shell of
particles. Loosely speaking, a particle ``has time'' for only one such
reaction while traveling a distance ${\cal O}(r)$.

The formation of a chromosphere seems unlikely on purely phenomenological
grounds. Note that the number of interactions per parton is proportional to
the number of initial partons. If ten initial partons lead to a chromosphere,
five or six initial partons should at least show significant effects from
these interactions. The observation of events with six well--defined jets has
been reported by both the UA2 \cite{ua2} and CDF \cite{cdf} collaborations.
CDF finds fair agreement between observations and QCD parton shower
simulations (based on leading--order matrix elements). They demand that the
total invariant mass of the 6--jet system exceed 520 GeV, while each jet
should have a transverse momentum of at least 20 GeV. Using $N_{\rm init} =
6$, $Q = 30$ GeV and $\Gamma_{\rm bh} \longrightarrow M_{6-{\rm jet}} = 500$
GeV setting the scale for the initial hard reaction, Eq.(\ref{intrate})
predicts ${\cal N} \simeq 0.7 \ (9.5)$ interactions per parton for $Q_{\rm
  min} = 9 \ (2)$ GeV. We find it difficult to believe that such high
interaction rates would leave no imprint in the properties of the observed
events, given that 3 (30) interactions are supposed to lead to nearly complete
thermalization. 

Even if a nearly thermal chromosphere does not form, parton--parton scattering
might still have significant impact on the hadronic final state from black
hole decay. Given the complexity of QCD processes even in the absence of
parton--parton scattering, a quantitative investigation of their effect can
only be performed with the help of a QCD simulation program. This is the topic
of the next Section.

\section{Simulation}

We saw in the previous Section that a quantitative analysis of the effects of
parton--parton scattering is only possible if we also treat the QCD parton
showers that occur whenever a large four--momentum is transmitted to strongly
interacting particles. This implies that we need to follow the
spatio--temporal development of these QCD showers. In this Section we first
outline the general philosophy of our approach; a more detailed description
of the various stages treated by our code will be given in the subsequent
Subsections. 

Numerical codes that simulate parton showers are probabilistic, i.e. they
operate with squared amplitudes. Quantum mechanical interference effects can
therefore only be treated approximately (e.g. through angular ordering
\cite{order}, whereby subsequent gluons are emitted at smaller and smaller
angles.) The basic idea is that partons emerging from a hard reaction
(scattering or decay) initially have time--like virtuality, which is reduced
by ``branching off'' additional partons. This can be justified by the
observation that in a complete calculation of the relevant Feynman diagrams,
final state partons emitting additional partons indeed have to have time--like
virtualities. The beauty of such showering algorithms is that they exploit QCD
factorization theorems to sum such higher order processes to all orders in
perturbation theory, albeit (usually) only with leading logarithmic accuracy.

Most of these codes follow the ``evolution'' of this shower not in time, but
in an energy variable, which in the simplest case is given by the virtuality
of the partons in the shower; since we are dealing with final--state showers,
the partons in question have time--like momenta, i.e. the shower has the same
kinematics as a cascade of two--body decays. (In this analogy, the lifetime of
the decaying particles would be given by the inverse virtuality of the
particles in the shower.)

According to quantum mechanics, one cannot simultaneously determine this
shower energy scale and the (proper) time of a branching. Unfortunately for
our application we need to do precisely that. A certain additional abuse of
the principles of quantum mechanics is therefore inevitable. We do this by
identifying the {\em uncertainty} in time, as determined by the uncertainty
principle, with the actual {\em duration} of a given process. We use this
identification both for the branching and for parton--parton scattering; in
the former case, the time is given by the inverse of the virtuality of the
branching parton, whereas the time needed for a scattering is given by the
(space--like) virtuality of the parton exchanged in this scattering reaction.
Note that neither two branching steps, nor two scatterings, involving the same
partons can occur at the same time. In the absence of scattering, the
evolution in time would therefore strictly match the evolution towards smaller
virtualities. Moreover, between branching or scattering events, the partons
are taken to propagate along their classical paths. Note that this evolution
is nonetheless probabilistic, since after each branching or scattering the
4--momenta of the outgoing partons are chosen randomly, with distributions
determined by perturbative QCD (and subject to energy--momentum conservation).

We only aim at leading logarithmic accuracy. This means that we only use
leading order cross sections, and only include $2 \rightarrow 2$ scattering
processes; later gluon emission, which is treated explicitly in the estimates
of interaction rates in ref.\cite{Anchordoqui:2002cp}, is taken into account
by the subsequent shower evolution. In fact, including $2 \rightarrow 3$
processes in the scattering reactions would lead to double counting.
Scattering reactions can nevertheless increase the particle multiplicities,
since they can {\em increase} the virtuality of the participating particles;
in contrast, each branching reduces this virtuality. Moreover, the scattering
can change the 3--momenta of the particles, thereby potentially destroying the
jet structure. As mentioned in the previous Section, one needs large
scattering angles in order to establish a chromosphere from an initially small
number of very energetic partons.

As usual, we treat showering and hard scattering independently, i.e. we apply
QCD factorization. This requires that the scattering indeed be sufficiently
hard; that is, the absolute value of the four--momentum exchanged in the
scattering reaction must be (much) larger than the virtualities of the partons
in both the initial and final state. However, as already noted, the partons in
the final state may be more off--shell than those in the initial state. Note
also that we should not include initial state showering in our approach, since
each ``initial state'' of a scattering reaction is part of the extended final
state shower that follows the Hawking evaporation of the black hole; including
initial state radiation would therefore also lead to double counting.

Our simulation starts with a rather small number of energetic, and far
off--shell, partons, using the results of Section~2. It then uses small time
steps to follow the parton shower and/or scattering of all partons. This phase
ends when all partons are (nearly) on--shell and far apart from each
other, so that neither further branching nor further scattering is possible.
When the particles virtualities reach the QCD scale $\Lambda_{QCD}$,
hadronization will take place and the data of the final particles will be
stored for statistical analysis.

The simulation code has been written nearly from scratch in C++, although the
global structure is based on the VNI 4.12b Monte Carlo simulation
\cite{Geiger:1998fq, Bass:1999pv} using its particle record and a selection of
modified routines from it. The VNI particle record uses the ``Les Houches''
format, extended to hold information necessary for the full space--time
evolution of the partons. VNI 4.12b was originally written in order to
simulate ultra--relativistic heavy--ion collisions; in that case, multiple
partonic scattering reactions are certainly important, and had to be modeled
carefully, making this code a good starting point for our work. In contrast,
scattering in the final state is thought to be unimportant for hard reactions
involving $e^\pm$ and/or $p/\bar p$ in the initial state.

We also use some routines from the PYTHIA Monte Carlo simulation
\cite{Sjostrand:2003wg}. For numerical integration and a simulation grade
random number generator the GNU Scientific Library \cite{gnu} for C/C++ was
used.

\subsection{Initial setup}
\setcounter{footnote}{0}

As in ref.\cite{Anchordoqui:2002cp} we randomly distribute the initial partons
inside a shell with thickness equal to the black hole lifetime $\tau_{bh}$
(\ref{life}) around the decayed black hole with radius \eqref{radius}. Quarks
and gluons are generated separately, with average multiplicities given by
Eq.(\ref{nav}) and energy distributions according to Fig.~1. The momenta of
most partons are chosen randomly inside that half of the solid angle which
points away from the black hole, as seen from the location at which the
particle is created. Choosing the partons to be always emitted radially by the
black hole, which would be proper for non--rotating black hole, would make
future collisions between them impossible in the absence of showering; our
choice therefore increases the possibility of parton--parton
scattering.\footnote{Non--radial emission should occur during the spin--down
  phase, with quite complicated angular dependence
  \cite{Ida1,kanti,Ida:2006tf}. Since most of the energy is released after the
  black hole has shed its spin, our treatment most likely over--estimates the
  probability of parton--parton scattering.} The 3--momentum of the last
parton is taken such that the total 3--momentum in the black hole rest frame
is zero.\footnote{This is not always possible, given that the energy of this
  parton has already been fixed and its virtuality must be smaller than this
  energy. If no solution is found, the event is abandoned, and the routine for
  generating the initial set--up makes a new attempt.} This is not absolutely
necessary, because the 3--momenta of the strongly interacting particles alone
do not have to add up to zero, but it allows a good control of the simulation.

Next, quark flavors are assigned randomly. We do not include the top and
bottom quark because we cannot assume them to be massless in all possible
collisions when we are using the massless QCD scattering amplitudes; indeed,
top production is likely to be somewhat suppressed since for our choice of
parameters, the Hawking temperature is somewhat below $m_t$. Charge
conservation is not enforced in our initial setup taking into account that the
black hole also radiates off other particles like leptons. However we take
care that the charges add up to some integer.

We set the maximal initial virtualities of the partons equal to their
energies; the actual values of these virtualities will be chosen by the shower
algorithm described in a subsequent Subsection. This choice of maximal
virtuality reproduces features of hadronic $Z$ decays fairly accurately, when
starting from leading order, $Z \rightarrow q \bar q$, decays.\footnote{One
  might argue that this choice is not appropriate for black hole decay, since
  black holes are not pointlike, unlike $Z$ bosons. However, for choices of
  $M_D$ and $M$ relevant for LHC phenomenology, the scales $1/r_h$ or
  $\Gamma_{bh}$ that describe the spatio--temporal extension of the black hole
  are quite close to the average energies of the decay partons; note also that
  the final results will only depend logarithmically on the maximal initial
  virtuality.}

After the initial setup the program enters the main loop, which simulates the
space--time evolution of the parton cascade. This evolution is determined by
two processes, parton scattering and branching, which are described in the
next two Subsections.


\subsection{Parton scattering}
\setcounter{footnote}{0}

Every possible pair of partons is boosted from the black hole rest frame into
its cms frame, and it is checked if it has reached its closest possible
distance. In the next step it is checked whether partons which reached their
closest distance undergo a collision. We are using the cascade approach
\cite{Geiger:1991nj} for this purpose, according to which a collision takes
place if the closest distance of the parton pair is within the radius defined
by the total cross section of the specific process,
\begin{equation} \label{collision}
|\mathbf r_{a} - \mathbf r_{b}|_{\rm min} \leq \sqrt{ \frac
 {\hat{\sigma}_{ab}} {\pi}} \, .
\end{equation}

If there is more than one possible scattering channel for two partons the
total cross section will be the sum of the cross sections for all possible
final states,
\begin{equation} \label{sumcross}
\hat{\sigma}_{ab} = \sum_{c,d} \hat{\sigma}_{ab \rightarrow cd} \, .
\end{equation}
The individual cross sections are calculated numerically by integrating the
corresponding differential cross sections
\begin{equation} \label{diffcross}
\hat{\sigma}_{ab \rightarrow cd} = \int_{\hat{t}_{\rm min}}^{\hat{t}_{\rm max}}
\left(\frac{d\hat{\sigma}(\hat{s}, \hat{t}, \hat{u})} {d\hat{t}} \right)_{ab
  \rightarrow cd} d\hat{t} \, ,
\end{equation}
with Mandelstam variables $\hat{s} = (p_a + p_b)^2$, $\hat{t} = (p_a -
p_c)^2$, $\hat{u} = (p_a - p_d)^2$. The choice of the upper and lower bounds
$\hat{t}_{\rm max}$, $\hat{t}_{\rm min}$ in equation \eqref{diffcross}
requires some care. When two on--shell partons are scattering, the matrix
elements for the ($2 \rightarrow 2$) QCD cross sections diverge for forward
($\hat{t} \rightarrow 0$) and/or backward ($\hat{u} \rightarrow 0$)
scattering, thus a minimal momentum transfer is needed that determines
$\hat{t}_{\rm max}$ and $\hat{t}_{\rm min}$ for given $\hat{s}$. In this case
we take the commonly used value (in the parton--parton cms) of $p_{\perp {\rm
    min}} = 1$ GeV; this requires $\sqrt{\hat{s}} > 2$ GeV. Collisions with
$p_\perp < 1$ GeV are considered to be soft and are not evaluated since only
collisions generating high transverse momentum can change the jet structure of
the event.

Collisions with at least one virtual particle in the initial state need
special treatment, since off--shell initial or final partons lead to non gauge
invariant ($2 \rightarrow 2$) amplitudes \cite{Geiger:1991nj}. The authors of
\cite{Geiger:1991nj} solve this problem by combining the scattering with
space-- or time--like branching in a single (rather large) time step so that
at the end of a scattering only on--shell particles are left and only
on--shell particles will scatter again. This should be sufficient for the
simulation of heavy ion collisions, where (nearly) all virtualities and
exchanged 4--momenta are quite small. However, in our case this procedure
would allow a parton to instantaneously\footnote{Our program uses much shorter
  time steps than the original VNI code, typically $\sim 2 \cdot 10^{-4}$
  GeV$^{-1}$ initially. We checked that choosing even shorter steps does not
  change the result.} shower off a virtuality of hundreds of
GeV and then undergo a collision with momentum exchange of only a few GeV.
This does not make sense, since such a relatively soft collision would take
much more time than the (initial part of) the showering; moreover, the (many)
particles produced in the shower might undergo scatterings of their own.

We therefore take another approach: we allow virtual particles to take part in
a collision only if the scattering scale $Q^2_{\rm scatt}$ is at least as high
as half the sum of the virtualities $Q^2_a$, $Q^2_b$ of the particles $a$ and
$b$ in the initial state,
\begin{equation} \label{qscale}
Q^2_{\rm scatt} \geq \frac{Q^2_a + Q^2_b}{2} \, .
\end{equation}
In this case the scattering will at least not take longer than the parton
shower up to that point. Moreover, it can be hoped that the scattering is hard
enough that the virtuality of the particles in the initial state becomes
irrelevant, so that we can describe the scattering using massless ($2
\rightarrow 2$) QCD cross sections. In fact, basically the same condition is
chosen by the usual QCD simulators when setting the showering scale for
initial state radiation, although in these programs the scattering scale
$Q^2_{\rm scatt}$ is fixed first. In our simulation the scattering scale is
taken to be
\begin{equation} \label{qscatt}
Q^2_{\rm scatt} = \frac {\hat{t} \hat{u}} {\hat{s}} 
\simeq \mathbf p_{\perp}^2 \, .
\end{equation}
This also fixes the scale in the running strong coupling constant,
\begin{equation} \label{alphas}
\alpha_s(Q^2_{\rm scatt}) = \frac{12\pi}{25\ \log(\frac{Q_{\rm scatt}^2}
  {\Lambda_{QCD}^2})} \, ,
\end{equation}
where we took $N_F = 4$ active flavors and QCD scale $\Lambda_{QCD} = 0.2$ GeV.
The requirement \eqref{qscale} determines the boundaries of the
integration over $\hat{t}$ in \eqref{diffcross}:
\begin{equation} \label{tbounds}
\hat{t}_{\rm min, \, max} = \frac{1}{2} \left( -\hat{s} \mp \sqrt{\hat{s}^2 - 
    4Q^2_{ab}\hat{s}} \right) \, ,
\end{equation}
where $Q^2_{ab} = \frac{Q^2_a + Q^2_b}{2}$. 

We did not include ($2 \rightarrow 1$) fusion processes like $g + g
\rightarrow g^*$ (with $g^*$ being off--shell), since the first branching of
the produced off--shell parton would again result in a $2 \rightarrow 2$
process, leading to double counting. It could be argued that we should also
require $Q_{\rm scatt} \geq 1 / |\mathbf r_{a} - \mathbf r_{b}|_{\rm min}$.
However, this would exclude interactions between partons that happen somewhat
before or after the time of their closest approach, which seems unphysical. We
therefore do not impose this additional requirement, which would reduce the
number of partonic scatterings significantly. In fact, the opposite
requirement, $Q_{\rm scatt} < 1 / |\mathbf r_{a} - \mathbf r_{b}|_{\rm min}$,
seems more reasonable, since according to the uncertainty principle a highly
virtual exchanged parton can only travel a very short distance. We do not
impose this constraint, i.e. we allow very large momentum exchange also
between relatively distant partons. This again increases the importance of
collisions, since a large $Q_{\rm scatt}$ is required for a collision to
modify the jet structure. However, we will see that not imposing this upper
bound on $Q_{\rm scatt}$ has very little effect in practice, since high values
of $Q_{\rm scatt}$ are in any case very unlikely.

In some rare cases one parton has two possible collision partners in a single
time step. In that case for both partners cross section and distance will be
checked. If only one collision takes place this will be the one evaluated. If
both collisions would take place only the one with the larger cross section
will be evaluated.

After a pair of particles has been identified as colliding, the scattering
kinematics can be generated. For initial states which have more than one
channel, the final state is chosen according to the probability given by the
relevant total cross sections. Next the value of $\hat{t} \in \left[
  \hat{t}_{\rm min}, \, \hat{t}_{\rm max} \right]$ is generated, with
distribution given by the corresponding differential cross section $d \hat
\sigma/ d \hat t$ (normalized to unity). The initial particle pair is boosted
into its cms frame and put on--shell. The absolute value of the transverse
momentum of the final particles is given by $Q^2_{\rm scatt}$ as determined by
the chosen value of $\hat{t}$. The azimuthal scattering angle, which fixes the
direction of the $\mathbf{p_\perp}$ vectors, is chosen randomly, with flat
distribution between 0 and 2$\pi$.

Finally the virtualities of the outgoing particles are created by the
branching routine (see the next Subsection) and color charges are assigned
according to the color flow of the specific process. It should be noted that
in case of the scattering of virtual particles, the virtuality after
scattering can become even lower than before, since $Q_{\rm scatt}$ is the
{\em maximal} virtuality of the particles in the final state. We see no
physical reason why such virtuality--reducing scattering reactions should be
suppressed. 

Due to the lifetime of a virtual (exchanged) particle the scattering will take
a finite amount of time determined by the scattering scale. In the rest frame
of the black hole this time is given by
\begin{equation} \label{tscatt}
\tau_{\rm scatt} = \frac{\gamma_{ab}}{Q_{\rm scatt}} \, ,
\end{equation}
where $\gamma_{ab} = (E_a + E_b) / \sqrt{\hat{s}_{ab}}$ is the boost factor
between the cms of colliding partons $a$ and $b$ and the black hole rest
frame. During this time the colliding particles are not allowed to collide
again, nor can they branch off additional partons; in fact, in quantum
mechanics one cannot say whether the parent partons $a,b$ or the children
$c,d$ exist during the period $\tau_{\rm scatt}$. After the scattering time is
over the partons are able to initiate final state radiation or collide
again.\footnote{In principle it might be more appropriate to place this ``dead
  time'' symmetrically around the time of closest approach, rather than
  letting it start at the time of closest approach. However, this would be
  technically difficult, since the program would then have to go back in time,
  and ``un--do'' any possible branchings that happened in the ``dead'' period
  before the time of closest approach. Since by construction $Q_{\rm scatt}$
  is larger than the virtuality of the scattering particles, this asymmetric
  placement of the ``dead time'' should not matter much in practice.}


\subsection{Parton branching}
\setcounter{footnote}{0}

While the inclusion of partonic collisions is the main new ingredient of our
simulation, we will see that parton branching plays a far more important role
in determining the characteristics of the final state. We model this using a
modified version of the relevant routine of VNI 4.12b, which in turn is based
on the PYTHIA branching algorithm \cite{Sjostrand:2003wg, Bengtsson:1986et,
  Webber:1986mc}.

The modifications implemented by us address the need to do branching for a
single parton, instead of treating two partons at once as is done normally in
order to ensure 4--momentum conservation. To make sure we can conserve energy
and momentum for a single parton we always have to determine the virtualities
at which the parton branches one step ahead in the branching algorithm: the
kinematics of $a \rightarrow b + c$ can only be fixed if the ``masses'', i.e.
(time--like) virtualities, of all three participating partons are known. Our
version of the routine therefore determines the virtualities of $b$ and $c$
earlier than the original routine does. Note that these actual virtualities
are typically much smaller than the corresponding maximal values; this is the
reason why energetic partons, with initial {\em maximal} virtualities given by
their energies, typically produce quite narrow jets. Another modification is
the introduction of the scattering--induced ``dead time'' described at the end
of the previous Subsection, during which partons are not allowed to branch.

Just like scattering reactions, the branching $a \rightarrow b + c$ also takes
a finite amount of time in our simulation, given by
\begin{equation} \label{tbranch}
\tau_{\rm branch} = \frac{\gamma_a}{Q_a} = \frac{E_a}{Q_a^2} \, ,
\end{equation}
with $\gamma_a = E_a/Q_a$ describing the boost from the rest frame of parton
$a$ to that of the black hole. In our simulation this is treated like the
lifetime of a decaying particle, i.e. the probability of an ``active'' parton
$a$ to branch during the time step $dt$ is given by \cite{Geiger:1998fq,
  Bass:1999pv}
\begin{equation} \label{pbranch}
P_{\rm branch} = 1 - {\rm e}^{-dt/\tau_{\rm  branch}} \simeq \frac {dt}
    {\tau_{\rm branch}} \, ,
\end{equation}
where the second, approximate equality holds for the (realistic) situation $dt
\ll \tau_{\rm branch}$. Note that each parton is checked for possible
scattering before it is checked for branching. In our treatment partons $b,c$
are created instantaneously if a branching occurs, i.e. they are allowed to
undergo scattering immediately.\footnote{These partons are created at slightly
  different locations, as determined by the uncertainty principle; moreover,
  they are moving away from each other. These two partons can therefore not
  scatter on each other, even if they are very close, but they can scatter on
  other partons.} This can be justified from the requirement (\ref{qscale}),
which ensures that scattering reactions are much faster than branchings.
Imposing a branching ``dead time'' of (some fraction of) $\tau_{\rm branch}$
on $b,c$ before they are allowed to scatter would obviously reduce the number
of parton--parton scatterings, although numerically this effect is not very
large. The 4--momenta of partons $b,c$ are chosen as in the PYTHIA branching
algorithm \cite{Sjostrand:2003wg}.

There are three situations in which the branching routine is used. First, all
initial partons which have some maximum virtuality from black hole decay are
sent through the branching routine once, in order to determine their actual
starting virtuality for the simulation. Secondly, the normal use when every
time step all partons which have any time--like virtuality left, and which did
not scatter recently, are sent through the routine to undergo branching with
probability given by Eq.(\ref{pbranch}). Third, similar to the first case, the
two partons coming out of a hard scattering which took place at the scale
$Q^2_{\rm scatt}$ are sent through the branching routine to determine the
actual virtualities they will start to propagate with.

The next step after the parton branching algorithm is to propagate all partons
freely by one time step. Then the simulation calculates the average distance
between all pairs of partons to see if it exceeds the QCD scale
$1/\Lambda_{QCD}$, at which point the partonic simulation loop can be aborted.
If the abort condition is met hadronization will take over, or else time is
increased by one time step and the simulation loop starts again. For details of
the hadronization algorithm we again refer to the documentation of PYTHIA
\cite{Sjostrand:2003wg}.


\section{Results}

We are now ready to present some numerical results. As stated in Sec.~2, we
set the higher--dimensional Planck scale to its lower bound, $M_D = 0.65$ TeV,
since this maximizes the number of initial partons, and hence the number of
parton--parton interactions within a given black hole decay. We also consider
relatively heavy black holes, with $M = 5$ and 10 TeV. Recall that decays of
heavier black holes are characterized by lower temperatures, and hence higher
multiplicities; however, the production cross section for black holes with
mass exceeding 10 TeV is certainly negligible at the LHC.

Since the initial set--up is created totally randomly there are large
event--to--event fluctuations of the number and energies of the initial
partons. One therefore needs sufficient statistics to make reliable statements
about average quantities; the results presented below are based on 200 events
for each run. In order to illustrate the effects of parton--parton
interactions, we made separate runs for each black hole mass where these
interactions were turned on or turned off. In the latter case the
spatio--temporal evolution of the parton shower is irrelevant, i.e. our
program reproduces standard (PYTHIA) showering for the given set--up of
original partons.

If black hole decays led to formation of a chromosphere, observables based on
jets would no longer be useful. In the following Subsections we therefore
analyze the distribution of two observables whose definition does not assume
the existence of jets: angular correlations between energetic charged hadrons,
and the overall energy flow. Of course, the observables are chosen such that
their distribution would be greatly affected if parton--parton scattering did
indeed lead to formation of a chromosphere. In the last Subsection we will
interpret these results with the help of the time structure of the parton
shower that develops after typical black hole decays.


\subsection{Angular correlation}
\setcounter{footnote}{0}

After hadronization we are left with a large number of charged, long--lived
particles (mostly charged pions and kaons and some protons) as well as photons
and neutral hadrons. In this first analysis we focus on charged particles
because their momenta can be measured accurately using tracking information.
We only consider particles with an energy (in the black hole rest frame) above
4 GeV; the number of charged particles passing this cut is denoted by $N_{\rm
  ch}$. This cut should largely remove particles from the underlying event
which have nothing to do with black hole decay, and which have not been
included in our simulation.

We compute the angles $\theta$ between any two of these charged particles;
altogether there are $N_{\rm pair} = N_{\rm ch}(N_{\rm ch} - 1) / 2$ such
pairs. The resulting distribution is binned in $\cos\theta$. Because $N_{\rm
  ch}$ can vary a lot from event to event, we normalize the distribution for
each event to $N_{\rm pair}$, before averaging over the 200 generated events.

Let us first consider some simple situations, in order to get a feeling for
what kind of distribution to expect. The simplest case, which certainly does
not describe black hole decay, would be events with two back--to--back jets,
each of which contains the same number $N_{\rm ch}/2$ of charged particles. In
this case $N^2_{\rm ch} / 4 - N_{\rm ch} / 2$ pairs would have an angle near
zero between the particles of the pair because they reside inside the same
jet; the remaining $N^2_{\rm ch} / 4$ pairs would have an angle near $\pi$
between the particles of the pair because the particles are in different jets.
In the limit of large $N_{\rm ch}$ these numbers will become almost equal,
leading to peaks at $1$ and $-1$ in our plot which have approximately the same
height.

A case that is closer to what one may expect from the decay of a black hole
would be an event with $n_j$ jets. Let us keep the assumption that each of
these jets contains exactly the same number $N_{\rm ch}/n_j$ of energetic
charged particles. In this case there are
\begin{equation} \label{pospairs}
\frac{N_{\rm ch}} {2} \left(\frac{N_{\rm ch}}{n_j} - 1 \right)
\end{equation}
pairs residing in the same jet; the remaining $N^2_{\rm ch} (n_j -
  1) / (2n_j)$ pairs of charged particles are in different jets. We thus see
that the fraction of all pairs that reside inside the same jet will decrease
when the number of jets increases, i.e. the peak in our distribution at
$\cos\theta = +1$ will become smaller. Moreover, for $n_j \gg 2$ randomly
distributed jets the peak at $\cos\theta = -1$ will vanish because momentum
conservation no longer requires any two jets in the event to be
back--to--back. After averaging over many events all the angles between
different jets contribute more or less equally. In this case we therefore
expect a smooth distribution (although momentum conservation my still lead to
a slight increase of the correlation function at negative $\cos\theta$) with
only one peak at $\cos\theta = +1$. The distribution should remain
qualitatively the same in the more realistic scenario where we allow different
jets to contain different numbers of charged particles, although
eq.(\ref{pospairs}) will then no longer be valid.

On the other hand, if we assume that a chromosphere is indeed spherical, the
correlation function should become almost perfectly flat; in particular, we
would not expect any visible peak at $\cos\theta = +1$.

Figs.~2 show results for black hole mass $M = 5$ TeV (top) and 10 TeV
(bottom), with (red) and without (black) parton--parton scattering.  Clearly
for both black hole masses there is still a strong peak at $\cos\theta = 1$,
leading to the conclusion that we should expect a jet structure after black
hole decay. The peak becomes smaller for a heavier black hole; this is
expected from the qualitative discussion presented above, given the increasing
initial parton multiplicity, see Eq.(\ref{nav}). The rise of the distribution
towards $\cos\theta=-1$ is also more pronounced for smaller black hole masses,
again as expected from our qualitative discussion. For $M=5$ TeV,
parton--parton scattering has essentially no effect on this distribution. For
$M=10$ TeV, it leads to a very slight broadening of the peak at
$\cos\theta=+1$; however, given the uncertainties of our simulation, we do not
claim that this effect is significant.

\begin{figure}[H]
\vspace*{-2cm}
\begin{center}
\includegraphics[scale=1.3]{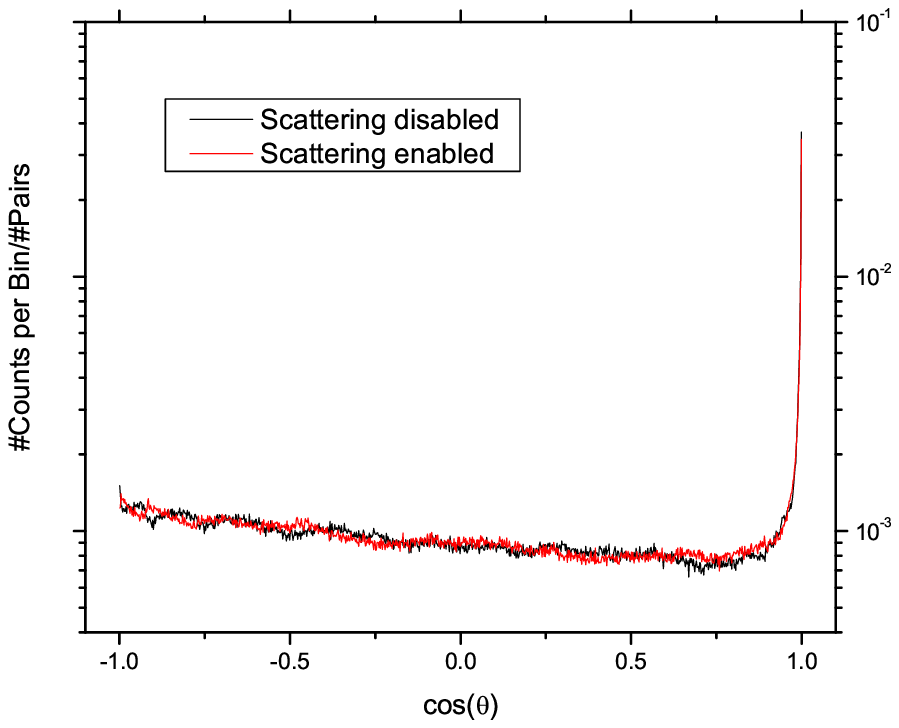}\\
\vspace*{-1.5cm}
\includegraphics[scale=1.3]{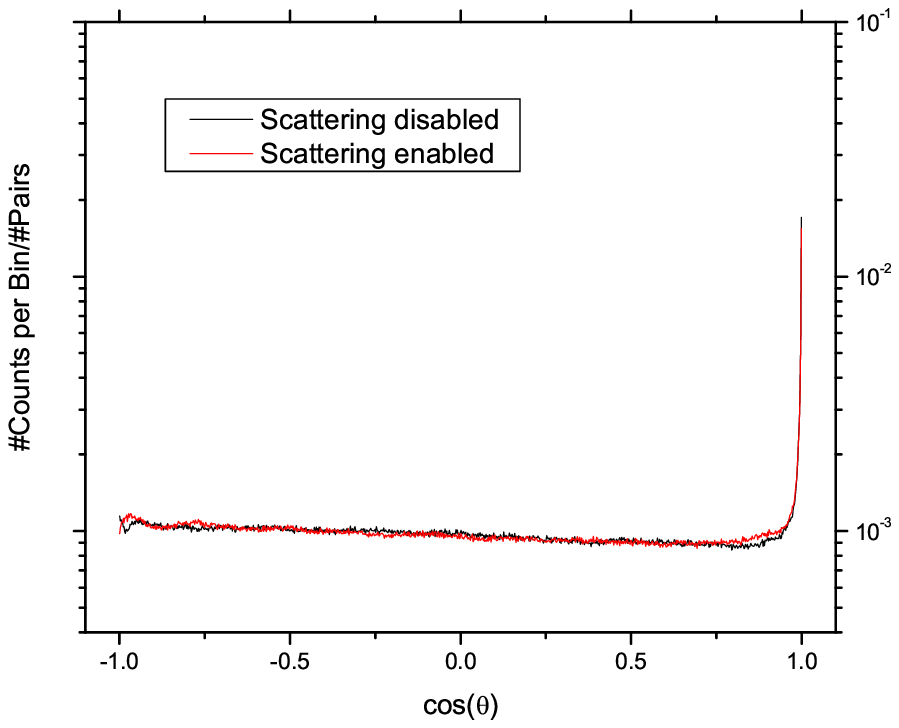}
\end{center}
\vspace*{-5mm}
\caption{The angular correlation function for charged particles from black
  hole decay with $E>4$ GeV, for black hole mass $M = 5$ (top) and 10 TeV
  (bottom), obtained by binning into 1,000 bins. The red and black curves have
  been obtained including and omitting parton--parton scattering,
  respectively.}
\end{figure} \label{fig_angle}


\subsection{Total energy flow}
\setcounter{footnote}{0}

The second quantity we investigate is the total energy flow from the hadronic
black hole decay products. Here we envision a calorimetric measurement. We
therefore divide phase space into azimuthal angle $\phi \in [0, 2\pi]$ and
pseudo--rapidity $\eta \in [-4,4]$, with 15 bins in $\phi$ and 30 bins in
$\eta$. For each of these 450 ``calorimeter cells'' the total visible energy
is calculated, including hadrons and photons from hadronic decay, but no
neutrinos or muons. We present the result by plotting the number of cells with
deposited energy $E_{\rm cell} \leq E_{\rm max}$ as function of $E_{\rm max}$.

Let us again first discuss the possible shapes of this distribution for
different final parton configurations. If a chromosphere forms, we expect a
very large number of hadrons in the final state, each of which has a
relatively small energy. These would be distributed uniformly over phase
space, i.e. had we defined our cells as having constant length in $\cos\theta$
all cells would receive essentially the same energy. We prefer to use $\eta$
to parameterize the phase space, since the energy flow pattern will then be
invariant under motion of the black hole along the beam pipe. Since $d / d
\eta = \sin^2(\theta) d / d \cos\theta$, cells at small $\cos\theta$, i.e.
small $|\eta|$, will then receive significantly more energy than those at
large $|\eta|$. There should nevertheless be almost no empty cells; the
maximal energy deposit, in cells with $\eta \sim 0$, would be about $0.009
E_{\rm tot}$, where $E_{\rm tot} \simeq 0.75 M$ is the total hadronic energy
released in the decay of a black hole with mass M. In particular, there should
not be any cells with energy comparable to the initial average partonic
energy, which amounts to about 300 GeV for our choices of parameters [see the
discussion of Eq.(\ref{nav}) in Sec.~2]. The distribution expected for
``ideal'' chromospheres, with energy flow being completely independent of
$\phi$ and $\cos\theta$, is depicted by the blue step--like\footnote{These
  steps appear only for black hole decays at rest. In general one expects a
  smoothed-out version of these curves once a distribution of longitudinal
  momenta of the black holes is taken into account.} curves in Figs.~3.

In the opposite extreme, where the final state consists of a relatively small
number [given by Eq.(\ref{nav})] of very narrow jets, most cells would be
empty, while in a few cells the deposited energy would be of order 300 GeV.
However, even in the absence of parton--parton scattering, final state
radiation implies that many jets will spread out over several cells. Together
with the final hadronization step, this will lead to a significant number of
cells in which a small, but nonzero, amount of energy is deposited, often in
form of a single hadron. We therefore expect a non--trivial dependence on
$E_{\rm max}$ in the entire range between $\sim 100$ MeV and 1 TeV.

Our results for the energy flow of hadronic black hole decay products are
shown in Figs.~3. We see that the number of cells with $E < E_{\rm max}$
indeed shows nontrivial dependence on $E_{\rm max}$ over a wide range. The
number of (almost) empty cells decreases with increasing black hole mass, as
expected from the higher initial parton multiplicity (\ref{nav}).  However,
even for $M=10$ TeV, in about 50\% of the cells almost no energy is deposited;
on the other hand, about twenty cells contain more than 100 GeV. We saw above
that these results are consistent with the existence of well--defined jets.
Parton--parton scattering has practically no effect on the number of cells
containing at least 30 GeV, again indicating that it does not affect the jet
structure at all.

\begin{figure}[H]
\begin{center}
\vspace*{-2cm}
\includegraphics[scale=1.3]{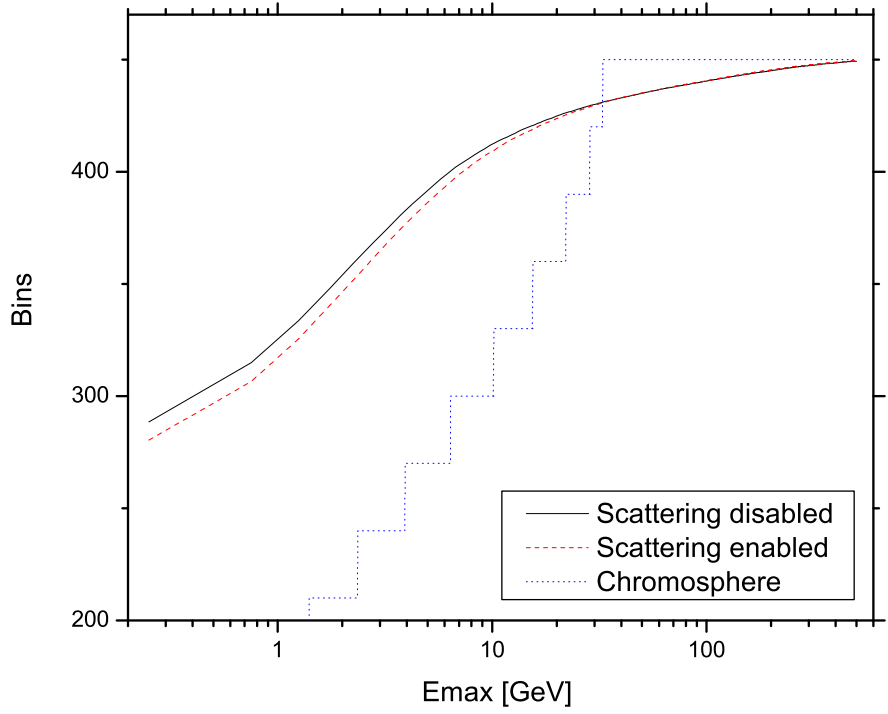}\\
\vspace*{-1.5cm}
\includegraphics[scale=1.3]{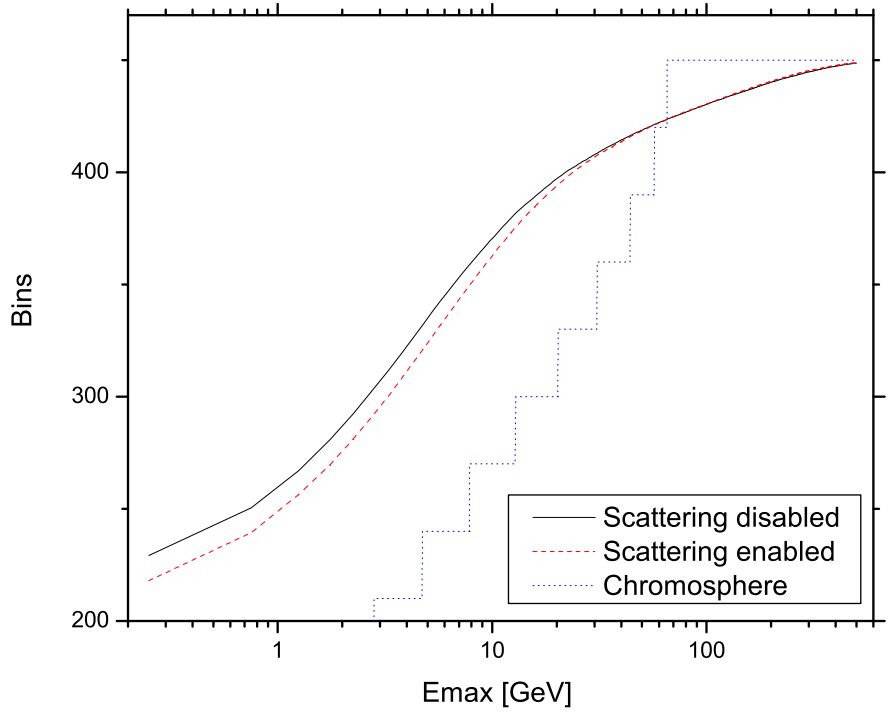}
\end{center}
\vspace*{-5mm}
\caption{The number of phase space cells where the deposited energy satisfies
  $E < E_{\rm max}$ as function of $E_{\rm max}$, for black hole mass $M = 5$
  TeV (top) and 10 TeV (bottom). The notation is as in Fig.~2, except that the
  blue, step--like, curves show the prediction from an ideal chromosphere.}
\end{figure} \label{fig_energy}

Scattering does increase the total number of non--empty cells by about 5\%
(10\%) for $M=5$ (10) TeV. However, the production of soft hadrons is a
non--perturbative process, and can therefore not be treated from first
principles; it is not clear whether the effect of parton--parton scattering is
larger than the systematic error of our simulation.\footnote{The {\em
    relative} size of the effect of parton scattering should be rather
  insensitive to the details of the simulation; the statement that it
  increases the number of non--empty cells by 5 to 10\% should therefore be
  relatively robust. However, at the end one can only compare the {\em
    absolute} prediction of the simulation with actual events.} Moreover, it
should be kept in mind that the ``underlying event'', which is created by the
remnants of the colliding protons that do not participate in black hole
formation, will also contribute a large number of (mostly soft) hadrons to the
final state, which have not been included in our simulation. It is therefore
not clear whether the increase of the cells containing some black hole decay
products results in a measurable difference in the total energy flow in the
event.


\subsection{Microscopic structure of the events}
\setcounter{footnote}{0}

The results of the two previous Subsections show that parton--parton
interactions have little effect on the global characteristics of the hadronic
final state that results from the decay of black holes with masses of a few
TeV. In this Subsection we analyze the microscopic structure of the evolution
of the hadronic final state that results from the decay of such black holes.

In Table~1 we list the average parton multiplicities (just before
hadronization), as well as the average number of parton--parton collisions,
for $M=5$ and 10 TeV. We see that even in the absence of parton--parton
scattering, QCD branching (final state radiation) increases the multiplicity
by an order of magnitude, relative to the initial multiplicity given by
Eq.(\ref{nav}). We also see that the average number of parton--parton
scatterings per black hole decay is not so small; as expected from simple
statistical arguments, it increases roughly quadratically with the (initial or
final) partonic multiplicity. Scattering increases the final parton
multiplicity by 16\% (23\%) for $M=5$ (10) TeV. This effect is even larger
than that on the number of non--empty calorimeter cells. Note, however, that
the number of scatterings still remains well below the number of partons even
for $M=10$ TeV.

\begin{table}[h]
\begin{center}
\begin{tabular}{|c|| c | c | c || c| c| c| c|}
\hline
$M$ [TeV] & \multicolumn{3}{c||}{w/o scattering } & 
\multicolumn{4}{c|}{with scattering } \\
& $\langle n_q \rangle$ &  $\langle n_g \rangle$ & $\langle n_{\rm parton}
\rangle$ &$\langle n_q \rangle$ &  $\langle n_g \rangle$ & $\langle n_{\rm
  parton} \rangle$ & $\langle n_{\rm scatt} \rangle$ \\
\hline
5  & 19 & 116 & 135 & 21 & 134 & 156 & 16 \\
10 & 41 & 241 & 282 & 50 & 295 & 346 & 53 \\
\hline
\end{tabular}
\end{center}
\caption{Average final quark, gluon and total parton multiplicities from the
  decay of black holes with 5 and 10 TeV mass, with and without including the
  effect of parton--parton scattering. The last column gives the average
  number of partonic collisions. The statistical errors on the average
  multiplicities are about 2 percent.}
\end{table}

On the other hand, the probability $\langle n_{\rm parton} \rangle / \langle
n_{\rm scatt} \rangle$ that a given parton resulted from scattering is not
negligible. It rises roughly linearly with the parton multiplicity, in
agreement with the estimate (\ref{intrate}). However, this expression, with
$Q_{\rm min} \sim 1$ GeV as in our simulation, greatly over--estimates the
number of scattering reactions even if we normalize it to the initial parton
multiplicity.

As mentioned earlier, parton--parton scattering can only destroy the jet
structure if it involves large momentum exchange, i.e. if the scale $Q_{\rm
  scatt}$ defined in Eq.(\ref{qscatt}) is large. In Fig.~4 we show a scatter
plot of $Q_{\rm scatt}$ values vs. time in the simulation; here entries from
100 decays of black holes with mass $M=5$ TeV have been collected. We see
that the average value of $Q_{\rm scatt}$ decreases with time, once $t > 1/
\langle E_{\rm parton} \rangle \sim 3 \cdot 10^{-2}$ GeV$^{-1}$. This can be
understood from the observation that at early times, most partons are quite
far off--shell; our condition (\ref{qscale}) then implies that early
scatterings must involve rather large momentum exchange. The existence of a
few early scatterings with low $Q_{\rm scatt}$ is due to the fact that the
initial virtualities with which the partons are actually created by the
simulation may be much smaller than their maximal values, which is given by
the energies of these partons.

\begin{figure}[H]
\begin{center}
\includegraphics[scale=1.2]{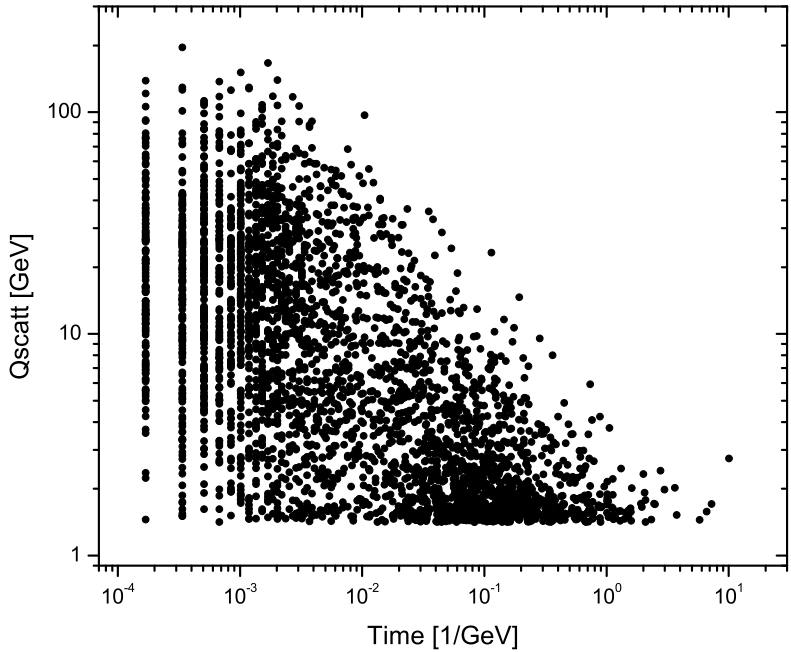}
\end{center}
\caption{Scatter plot showing the scale $Q_{\rm scale}$ defined in
  Eq.(\ref{qscatt}) vs. time after black hole decay, for $M=10$ TeV;
  scatterings from 100 decays have been added into this plot. The appearance
  of discrete lines at early times is an artefact of the finite time step size
  used.}
\end{figure} \label{fig_qscale}

Since the average virtualities of the partons diminish with increasing time,
more scatterings with small $Q_{\rm scatt}$ become possible. Since the QCD
cross sections satisfy $d \sigma / d Q^2_{\rm scatt} \propto \alpha_s(Q_{\rm
  scatt})^2 Q^{-4}_{\rm scatt}$ if $\hat t \hat u \ll \hat s^2$, reactions
with large $Q_{\rm scatt}$ are then greatly disfavored. Moreover, the average
parton energies also decrease with time. Later collisions therefore tend to
have smaller $\hat{s}$, which implies reduced kinematical upper bounds on
$Q_{\rm scatt}$.

The time dependence of the number of scattering reactions is determined by
several competing effects. While the number of partons increases due to
multiple branchings as time goes on, the number {\em density} of partons
decreases, making it increasingly less likely that two partons will come close
to each other. As a result, the number of collisions per unit time decreases
at first. On the other hand, the decreasing virtuality of the partons,
and corresponding decreasing lower bound on $Q_{\rm scatt}$, means that their
scattering cross section increases with time. This leads to an accumulation of
relatively soft scatterings at $t \sim 0.1/$GeV.

A typical decay of a 5 TeV black hole therefore develops as follows.
Initially there are about 12 highly off--shell partons with average energies
around 300 GeV. One or two of them may undergo scattering very soon after
their creation, with typical scattering scale $Q_{\rm scatt}$ of a few tens of
GeV.  Note that these rather hard reactions still mostly have $Q_{\rm scatt}
\ll E$, so that the direction of the participating partons is not changed very
much.  Moreover, these early scatterings may actually reduce the virtualities
of the participating partons. As time goes on, the number of partons in the
event increases, mostly due to branching processes. The number of scatterings
also increases, but most of these reactions are relatively soft. We saw in
Table~1 that they do increase the partonic multiplicity at the end of the QCD
cascade.  However, most of these scattering reactions involve partons inside
the same jet, and therefore have little effect on the global quantities we
analyzed in the previous Subsections. In fact, the creation of an additional
very soft parton nearby in phase--space may not be visible in the (hadronic)
final state at all. This explains why the results in Table~1 seem to indicate
larger effects from parton--parton scattering than what we saw in Figs.~2 and
3.

\section{Summary and conclusions}

In this paper we investigated the question if the decays of black holes that
might be produced at the LHC would be characterized by a relatively large
number of discrete jets, as is usually assumed, or if they would lead to the
formation of quasi--thermal ``chromospheres'' of rather soft hadrons, as
suggested in ref.\cite{Anchordoqui:2002cp}. To this end we first described the
partonic state produced by the decay of a black hole. We found that, if the
higher--dimensional Planck mass $M_D$ is at its current lower bound, the
average initial number of partons will be less than about 25 for black holes
with mass $M \leq 10$ TeV, which might have a chance to be produced with
appreciable rates at the LHC; higher values of $M_D$ would lead to initial
states with fewer partons, and hence to less scattering.

In Sec.~3 we summarized the argument in favor of formation of a
chromosphere. We pointed out that this argument ignores the fact that
scattering reactions take a finite amount of time. Moreover, a very similar
argument should apply to ordinary QCD multi--jet events. It would predict
sizable effects in six jet events, which have been studied experimentally by
the UA2 and CDF collaborations, who found good agreement with standard QCD
predictions which ignore interactions between the partons in the final state.

Clearly a dedicated Monte Carlo study is required in order to determine the
quantitative effects of such final state interactions on black hole events at
the LHC. As described in Sec.~4, we wrote such a simulation code, based on the
VNI program \cite{Geiger:1998fq, Bass:1999pv}. We modified it by forcing
scattering reactions to take a finite amount of time, which we estimated using
the uncertainty principle. A very similar argument implies that, as time goes
on, the parton system evolves towards smaller virtualities through branching
processes. Scattering reactions may increase the virtualities of the
participating partons again, but can also be a shortcut towards smaller
virtualities. 

Results of our simulation are described in Sec.~5. We found the effects of
parton--parton scattering in the final state to be essentially negligible both
for the angular correlation between energetic charged hadrons, and for the
number of phase space cells containing a large amount of energy. We found some
effect on the number of cells containing only one or a few soft hadrons. We
interpreted these results in terms of the microscopic structure of the event.
In particular, we saw that ``hard'' reactions, with large momentum exchange,
only occur early on; in this case they may well reduce the virtualities of the
participating partons. Later scatterings are all quite soft. At both early and
late times, most scattering reactions have momentum exchange well below the
energies of the participating partons; such reactions cannot significantly
change the directions into which these partons are traveling. We therefore
conclude that a chromosphere will {\em not} form; programs that ignore
interactions between the partons from the decay of black holes created at the
LHC only make a negligible mistake.

Our simulation has some shortcomings. Like all QCD simulation codes, it
essentially works on the level of squared matrix elements, i.e. quantum
mechanical interference effects can only be included in an approximate
manner. In addition, we had to identify the quantum mechanical uncertainty in
time with actual duration, both for the branching and scattering processes. It
is not clear to us how this limitation can be overcome, even in principle.

Our results starkly contradict the claims of ref.\cite{Anchordoqui:2002cp},
even though we chose our parameters and initial set--up such as to maximize
the likelihood of scattering reactions between partons in the final state. It
might therefore be worthwhile to summarize the three effects that have reduced
the number of such reactions in our simulation.

Perhaps most important is that we only allow scattering of partons that
approach each other. Interactions between partons that ``always'' (after their
creation) move away from each other can certainly not be treated using the
factorization into initial black hole decay and subsequent parton--parton
scattering that underlies our simulation; for one thing, no $S-$matrix could
be defined for such a state.\footnote{In some sense such partons can
  nevertheless still interact with each other. For example, the well--known
  Coulomb singularity \cite{schwinger} for near--threshold decays into charged
  particles can be interpreted in terms of the electromagnetic interaction
  between these particles. However, in the simplest case of two--body decays,
  this singularity only affects the partial width for the decay; it does not
  increase the final state multiplicity. In case of decays into more than two
  partons, it can also change kinematical distributions, e.g. the invariant
  mass distributions of subsets of partons. However, a proper treatment of
  these effects would require the inclusion of corrections to black hole
  decays due to gauge loops. It is currently not clear how this could be
  done.}

Strictly speaking, in our case one cannot define an $S-$matrix for partons
approaching each other, either, since they were created at a finite time, i.e.
they did not exist in an ``in-''state defined at time $t=-\infty$. We
circumvent this problem by requiring the 4--momentum exchanged in these
reactions to be larger than the initial virtualities. In this case one should
be able to approximately treat the scattering like that of on--shell
particles. This roughly means that we require scattering reactions to be fast
on the time scale of the shower evolution up to the time of the scattering.
This constraint greatly reduces the scattering cross section at early times,
when the parton density is highest, and therefore reduces the number of
scattering reactions. The fact that we do not allow partons participating in a
scattering to start another scattering while the first one is still ``in
progress'' has a much smaller effect on the number of these reactions.

Although the number of parton--parton collisions remains small for all black
holes that might be produced at the LHC, we saw that it increases roughly
quadratically with increasing black hole mass. Extrapolating from the results
of Table~1, we estimate that there will be ${\cal O}(1)$ parton--parton
scatterings per produced parton once $M \gsim 25$ TeV, if the
higher--dimensional Planck scale $M_D$ is kept at its lower bound of 0.65
TeV. This need not lead to formation of a chromosphere, however; we saw that
most scattering reactions have little effect on the jet structure of the event.
Note that in this case the average initial partonic multiplicity already
exceeds 100, making the reconstruction of distinct jets quite unlikely, even
if such heavy black holes could ever be produced at human--made colliders.

\subsubsection*{Acknowledgments}

The work of K.O. is partly supported by the Special Postdoctoral Researchers
Program at RIKEN.

\end{document}